# Learning to map source code to software vulnerability using code-as-a-graph

Sahil Suneja, Yunhui Zheng, Yufan Zhuang, Jim Laredo, Alessandro Morari
IBM Research, Yorktown Heights, NY

*Abstract*—We explore the applicability of Graph Neural Networks in learning the nuances of source code from a security perspective. Specifically, whether signatures of vulnerabilities in source code can be learned from its graph representation, in terms of relationships between nodes and edges. We create a pipeline we call AI4VA, which first encodes a sample source code into a Code Property Graph. The extracted graph is then vectorized in a manner which preserves its semantic information. A Gated Graph Neural Network is then trained using several such graphs to automatically extract templates differentiating the graph of a vulnerable sample from a healthy one. Our model outperforms static analyzers, classic machine learning, as well as CNN and RNN-based deep learning models on two of the three datasets we experiment with. We thus show that a code-as-graph encoding is more meaningful for vulnerability detection than existing code-as-photo and linear sequence encoding approaches.

## I. INTRODUCTION

In this paper, we explore the applicability of Graph Neural Networks in understanding source code nuances, from a security perspective. We focus on code written in the C programming language. Source code naturally lends itself to a graph model in several forms such as abstract syntax tree, control flow graph, data dependency graph, etc. We explore whether it is possible for a neural network to understand the relationships between the different *nodes* and *edges* of a source code graph, and further, if it is possible to automatically extract (*learn*) certain templates of these relationships. If this is possible, then using the principle of graph isomorphism, perhaps these templates can be matched across unseen source code graphs to uncover its properties.

We take the example of vulnerability detection in code to drive this study. Existing rule-based approaches to vulnerability detection, both in static and dynamic analysis, suffer from shortcomings which a learning-based approach may help alleviate (Section 2). In the deep learning domain, Graph neural networks [34] have been shown to be useful in tasks such as logic reasoning and program invariant inferencing [21], code categorization [38], and code variable prediction [11], which makes it intriguing to test their efficacy in source code level bug-finding. Furthermore, Yamaguchi et al. [39] have shown that manually-created vulnerability-specific graph-level templates are useful in searching for *similar* instances elsewhere in a project. Thus, we wish to combine the two approaches to see if neural networks can help automate the bug-finding process. This automated template-matching would then hopefully augment the existing static and dynamic analysis approaches, thereby resulting in a more effective secure code pipeline.

In this work, we perform supervised learning over labeled (buggy/clean) C functions. The idea is that since code has structure it can potentially be used to figure out if certain part of the code structure is related to specific vulnerability types. To this end, we first convert code to to a comprehensive graph representation called the Code Property Graph (CPG) [39]. Next, we encode the graph to a vector representation using Word2Vec [28] while preserving its semantic properties. The code graph vector is fed to a graph neural network, which learns the signature of bugs in terms of relationships between nodes and edges of the code's CPG.

Existing source-code-level learning-based approaches for vulnerability detection treat code as a picture or as a linear sequence of tokens. The *pixel signals* from such a picture are then extracted using Convolutional Neural Networks which have been shown to perform very well in the image processing domain [20]. Alternatively, temporal signals from the linear sequence encoding are extracted using Recurrent Neural Networks or their variants [23], [30]. We believe that a code-as-graph representation carries more meaningful signals that can be exploited for better vulnerability detection. For example, absence of a 'variable sanitization template' [39], i.e. value-range validation prior to being used as a memory allocation size argument, may indicate a potential buffer overflow vulnerability. This template can potentially be automatically learnt in terms of control-flow and data-dependency relationship between locations of interest in source code (i.e. *edge* connectivity between *nodes* of a *graph*). We explore this hypothesis in this work [1].

In addition to the comparison between the effectiveness of different source code encoding approaches in learning code-to-vulnerability mappings, we also share certain insights we gathered in this work. Specifically, as expected, classifying code into vulnerable or not turned out to be hard on real-world (noisy) data as opposed to synthetic (clean) data. Furthermore, per-project classifiers were much more accurate than a universal classifier across multiple real-world projects. Similarly, learning vulnerability-specific models was easier than a global model across all vulnerabilities.

We present quantitative evidence of these observations in SectionIV. We experimented with different datasets namely Draper [25], Juliet [29] and s-bAbI [36]. We used the Gated Graph Neural Network(GGNN) [21] to perform graph-level learning on the code snippets in these datasets. The GGNN model performed very well in the s-bAbI and Juliet datasets, achieving an F1 score of 0.99 and 0.87 respectively, and an average precision of 0.99 and 0.96 respectively. The results on the more complex Draper dataset was average with a 0.5 F1 score and 0.4 AP.

---
[1] There is another avenue of treating code-as-natural-language, where approaches such as BERT which have shown great promise in the NLP domain can be put to use. We are still exploring this dimension of code representation and do not compare with it in this work.

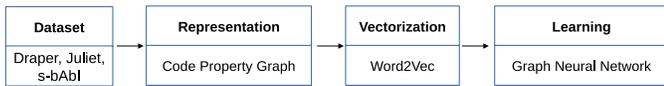

Fig. 1. AI4VA Pipeline

## II. Design

In this work, we focus at source code at a function-level. Given a C function the goal of our tool is to classify it as being vulnerable or not. We leave the task of localizing the bug for future work.

Different approaches have different ways of detecting vulnerabilities in code, such as (i) *rules* of a static analyzer, (ii) templates as in [39], and (iii) actual code execution and exploration as in fuzzers and dynamic analysis. It can be tedious and error-prone to have to create and maintain explicit rules and templates for different languages. That is perhaps why we see so many False Positives (code tagged as being vulnerable but isn't so) from static analyzers, causing developer fatigue in having to sift for True Positives (code tagged as being vulnerable and is indeed so) amongst numerous alerts of a static analyzer. On the other hand, having to compile and execute code and then hunt for specific inputs that can trigger a vulnerability, as in dynamic analysis, is very time consuming and lacks completeness.

In contrast, we wish to explore whether it is possible to learn such code-to-vulnerability mapping heuristics automatically. The idea is pretty straightforward–presenting a large dataset of tagged examples (vulnerable, not-vulnerable) to a learning-based model so that it can figure out the properties which differentiate vulnerable from healthy code. The learned heuristics from such a macroscopic approach to vulnerability detection can then augment the microscopic approach of static analyzers. The end result is a more finely curated set of alerts to the developer to aid in secure code development with hopefully lesser False Positives.

Figure 1 shows our AI4VA[2] pipeline which takes in raw source code and classifies it as being vulnerable or not. It consists of the following 4 stages:

### A. Dataset Acquisition

As in any learning-based study, access to a high quality dataset is quintessential. For our use-case we need a well-labeled source of C functions. Fortunately, a few of these are publicly available and we use 3 such datasets in our study: Draper [25], Juliet [29] and S-babi [36]. Out of these, S-babi and Juliet are synthetic datasets whereas Draper also

[2]Short for 'Artificial Intelligence For Vulnerability Analysis'

```
void foo()
{
   int a = 43
   char arr[55];
   if (a < 55)
   {
      a = 63;
   }
   arr[a] = 'X';
}
```

Fig. 2. A C function suffering from a buffer overflow vulnerability in the last line.

contains source code from the wild, specifically from Debian and Github projects. We test with multiple datasets to see how well a learning-based tool fairs when the dataset's signal vs. noise content varies.

### B. Code Representation

We do not perform an explicit feature extraction step in our AI4VA pipeline in contrast with classical Machine Learning (ML) approaches. This means we do not have to convert code into features such as number of lines, conditionals, identifiers and library-calls, cyclomatic complexity, Halstead complexity, call-stack depth, and the like. The Deep Learning models of today are capable of automatic feature extraction. So we focus on converting code into a representation amenable for consumption by Neural Networks, whilst preserving semantic information from the code.

Some existing approaches to code based vulnerability detection treat source code as photos, and then try to extract *pixel signals* from it using Convolutional Neural Network models from the image domain [20]. While others treat code as a sequence of tokens and try to extract temporal signals using Feedforward or Recurrent Neural Networks and its variants [23], [30]. In this work, we focus on the graph nature of source code which we feel is a more natural representation of code, encapsulating semantically rich information.

We use Code Property Graphs (CPGs) [39] to represent source code. A CPG is essentially a combination of the Abstract Syntax Tree (AST), the Control Flow Graph (CFG), and the Program Dependency Graph (PDG) extracted from source code. A C function is essentially converted into a graph containing different kind of nodes and edges conveying different kind of information. In particular, its syntax information is captured by the AST. For example, the nodes and the black (solid) edges in Figure 3 represent its AST. While AST models the structures of the function, it is insufficient to reason about program behavior without understanding the semantics of the structures on AST. To this end, semantic information is annotated atop AST, which represents the data flow and control flow information. For example, the blue (dash-dot) edge with tag "D_a" in Figure 3 shows the data dependency from the sub-tree that defines the variable a to the sub-tree that uses the defined value (a *Use-Def* edge). Similarly, the green (dashed) edges capture the execution order such as conditional branches. In this way, we can encode both syntax and semantics without losing any information.

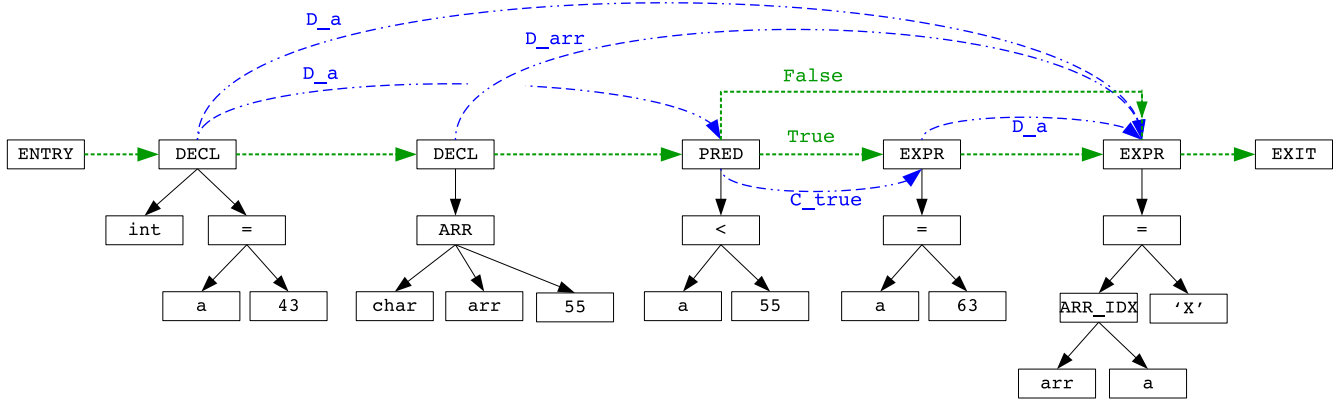

Fig. 3. A CPG for the code shown in Figure 2, simplified as per [39]. Edge-type legend: Black (solid) = AST, Green (dotted) = CFG, Blue (dash+dot) = PDG. Joern actually produces a more detailed CPG with almost twice as many nodes and edges as shown here.

| | NODES | |
|---|---|---|
| Node# | Type | Code |
| 3 | FunctionDef | foo () |
| 5 | IdentifierDeclStmt | int a = 43 |
| 6 | IdentifierDecl | a = 43 |
| 7 | IdentifierDeclType | int |
| 8 | Identifier | a |
| : | | |
| 17 | IfStatement | if ( a < 55 ) |
| 18 | Condition | a < 55 |
| 20 | Identifier | a |
| 21 | PrimaryExpr | 55 |
| 23 | ExpressionStmt | a = 63 |
| : | | |
| 36 | CFGEntryNode | ENTRY |
| 37 | CFGExitNode | EXIT |
| 39 | Symbol | a |

| | EDGES | |
|---|---|---|
| Node1 | Node2 | EdgeType |
| 6 | 7 | IS_AST_PARENT |
| 5 | 6 | IS_AST_PARENT |
| : | | |
| 36 | 5 | FLOWS_TO |
| 5 | 18 | REACHES a |
| 18 | 23 | CONTROLS |
| : | | |
| 5 | 39 | DEF |
| 18 | 39 | USE |
| : | | |
| 18 | 23 | DOM |
| 5 | 36 | POST_DOM |
| : | | |

Fig. 4. Output generated by Joern for the function shown in Figure 2. Shown is a list of CPG nodes and edges, indexed by node IDs.

| Target | Similar tokens |
|---|---|
| int | unsigned, long, char |
| 1 | 3,4,2 |
| + | -, /, +=, 2 |
| == | !=, !, >=, <= |
| if | &&, \|\|, while, -> |
| for | ++ |
| malloc | calloc, realloc, atoi |
| memset | memcpy, strncpy, copy_to_user |
| printf | fprintf, sprintf, strcpy, perror |

Fig. 5. Word2vec vector representation for C tokens preserves their semantic relationship. Shown here are tokens 'similar' to a target based upon Word2vec similarity $\geq 0.5$. As can be seen, Word2Vec picks up some weird nuances from the dataset such as the similarity between '+' and '2', and between 'malloc' and 'atoi'

We use the Joern open-source tool [17] to convert each sample from the dataset into its corresponding CPG. The CPG shown in Figure 3, for the code snippet from Figure 2, is a simplified version of the actual CPG which Joern generates. Figure 4 shows Joern's output in terms of a list of nodes and edges.

The set of edges for the CPGs in our target datasets consist of the following 12 edge types: IS_CLASS_OF, IS_FUNCTION_OF_AST, IS_FUNCTION_OF_CFG, IS_AST_PARENT, USE, DEF, DOM, POST_DOM, CONTROLS, DECLARES, FLOWS_TO and REACHES. We preserve the edge type information in the CPG representation of each sample of the dataset. As described in Stage 4, we chose a Neural Network model that respects and learns the properties and importance of these edge types in mapping code-to-vulnerability.

### C. Vectorization

This stage is necessary to convert human-readable CPG representation of code samples into a vector representation consumable by Neural Networks. An edge is converted to a tuple {*source_node_id, edge_type, dest_node_id*}. Tuples belonging to all edges in a CPG are then consumed by the graph neural network model (Stage 4) to create adjacency matrices per edge-type.

Each node is converted to a N-dimensional vector. A node can contain a variety of information depending upon its source- AST, CFG, PDG. It can contain C keywords (int, for), library function names (malloc, sprintf), operators (+, =), and identifiers ('a', 'foo'). In addition, it can also contain CPG *node type* identifiers such as *IdentifierDeclStatement, AssignmentExpression* etc. as shown in Figure 4. One approach is to randomly assign the nodes N-dimensional representations via one-hot encoding—a dimension for each token, represented with a '1' if the token exists in the node and '0' otherwise. Such random assignment results in loss of meaningful information. For example, the assigned vector representations of '0' and '1' might not reflect their integer *relationship*. To preserve these semantic attributes of the code and the language, we use Word2vec [28] to encode each token as an N-dimensional vector. As shown in Figure 5, inspired by [20], the vector representations of the digits is *closer* to each other, representing their semantic similarity, and distant from say that of memcpy and memset (themselves *closer* in encoding representing their memory allocation behavior similarity).

All identifiers are mapped to a generic 'ID' token, since they'll be different for different samples. An identifier `a` in a sample may not be the same `a` in another sample. We do not want to learn how the token named `a` relates to a vulnerability in the code, so we remove this noise by normalizing it to 'ID'. We still preserve it's signal by virtue of edge connections of this node to other CPG nodes.

Non-leaf nodes in a CPG are represented by the N-dimensional vector of their CPG *node type*. We discard the *code* attribute associated with such nodes since it's components eventually get repeated in the child nodes (see Figure 4). For leaf nodes, we experiment with two different node representations: (i) average, and (ii) concatenation of the vectors corresponding to the CPG *node type* and the individual tokens of the node's *code* attribute.

### D. Learning

We base our learning phase on the hypothesis that that since there exists an inherent structure to code, it should be possible to exploit it to automatically figure out if certain part of the code structure is related to specific vulnerability types.

By this stage, a C function has been converted to a CPG and appropriately vectorized. Furthermore, a good amount of structural and semantic information contained in the code has been preserved in the process. We are now ready to *learn* the abstract *signatures* of bugs or vulnerabilities in terms of relationships between nodes and edges of the code's CPG.

We use a state-of-the-art neural network that has the ability to operate on graph inputs- Gated Graph Neural Network [21]. It operates by transferring *information* from adjacent nodes over the connecting edges and aggregating such information at each node. Such exchange of information is repeated for a certain number of steps, with retention of old and assimilation of new information governed via Gated Recurrent Units- a successor to Recurrent Neural Networks. A final classification layer maps the information accumulated at each node to its individual confidence in signaling a vulnerable or healthy verdict for its parent CPG (or equivalently the code sample). A combination of confidence values across all nodes of a CPG is the final confidence value the model assigns to its vulnerability prediction per sample.

Each node's vector representation acts as the initial *information* it carries. A weighted sum of information flowing from its immediate neighbors is the new incoming information per node. During the process of training, the model learns the values of weights to assign to different nodes and edges, depending upon their contribution to information flow resulting in correct vulnerable or healthy signals across samples.

## III. EXPERIMENT SETUP

The goal of the experiments is to classify a given C function as being vulnerable or not. We compare the classification performance of our AI4VA pipeline with existing approaches, in terms of F1 and precision/recall scores: (i) Deep Learning Models: CNN and RNN from [31], Memory Networks from [14], (ii) A classical Machine Learning model–Bag-of-Words + RandomForest–from [31]. (iii) Static analyzers, with their scores borrowed from [31], [36]

*Testbed*: Experiments were performed on a 160 core 3.6GHz PowerNV 8335-GTB machine with 506GB RAM and 4 Tesla P100-SXM2-16GB GPUs. The software stack includes RHEL-7.6 host OS, NVIDIA driver 418.67, CUDA 10.1, python 3.6, and tensorflow 1.13.1.

### A. Datasets

Here we present some salient features of the datasets we experimented with. All datasets were run through the AI4VA pipeline separately. The train:validate:test split was 80:10:10.

1. *Juliet* : The Juliet Test Suite [29] contains synthetic code examples with vulnerabilities from 118 different Common Weakness Enumeration (CWE) [15]classes designed for testing static analyzers. From its 64K test cases we extract 200K functions, amongst which almost 25% are vulnerable. Samples tagged as 'bad' are labeled as 1, and the ones with a 'good' tags are labeled as 0.

2. *Draper* : The Draper dataset [25] consists of 1.27 million C/C++ functions across synthetic (Juliet, as above) as well as real-world Debian and Github projects. It has been labeled by running each code sample through multiple static analyzers, followed by manual verification, de-duplication, and categorization into 5 different vulnerability categories: CWE-119, CWE-120, CWE-469, CWE-476, and OTHERS (all other CWEs combined). It is a highly imbalanced dataset with only ¡10% samples being vulnerable. When not doing per-vulnerability training (Section IV-Csec-pervuln)), samples with any CWE tags are labeled as 1, rest as 0.

3. *s-bAbI* : The authors of s-bAbI [36] claim that the Juliet dataset is far too small and complex to use in learning to predict the labeled security defects. Their synthetic dataset contains syntactically-valid C programs with non-trivial control flow, focusing solely on the buffer overflow vulnerability (CWE-120). Although s-bAbI is simple as compared to real-world code, static analyzers exhibit poor recall on it. We used the s-bAbI generator to create a balanced dataset of almost 475K functions. Samples with a 'UNSAFE' tag are labeled as 1, and those with 'SAFE' as 0.

### B. GGNN Model Configuration

We used the GGNN implementation available at [27] to which we added a classification layer together with cross entropy loss function. In addition to the regular hyper-parameter tuning, we experimented with a variety of model configurations such as:

1) Calculating the graph's vulnerability verdict using (i) sum of confidences per node, (ii) average of confidences per node, and (iii) confidence values of a master per-graph node added to accumulate signals from all of the graph nodes

2) Learning node importance at a per-class level or in an absolute sense in terms of contributing to the overall graph verdict
3) Learning the importance of edges in terms of propagating discriminatory signals

Wherever necessary, measures were taken to tackle class imbalance such as:
1) Giving more weights to vulnerable class during loss minimization while training.
2) Using focal loss [24] instead of cross-entropy loss.
3) Ensuring batch shuffling maintains class ratio.
4) Training with more balanced data splits.

Results presented in the next Section are for the best performing configuration per dataset. The performance difference with different configurations is more incremental than extreme. The following model configuration worked across the board: node feature vector size = 32, learning rate = 0.001, batch size = 100000 nodes (max nodes per graph = 699), GRU hidden size = 100, number of GRU unrolling timesteps = 5, GRU activation function = tanh, optimizer = Adam, dropout keep_prob = 0.8.

## IV. EVALUATION RESULTS

We group our findings into following 3 categories:
1) Data Source: Real-world vs. Synthetic
2) Encoding: Code-as-graph vs. Others
3) Classifier's Vulnerability Scope: Generic vs. Specific

### A. Data Source: Real-world vs. Synthetic

Table 6 compares the performance of different vulnerability detection approaches on the three datasets. As can be seen, performance of the approaches drops across the board when going from a clean but narrow-scope synthetic dataset (s-bAbI) to a noisy but real-world dataset (Draper). Although the trend is as per expectations but the performance drop is rather extreme, going from almost perfect scores to as good as merely a coin-toss! For both s-bAbI and Juliet, GGNN beats the competition. Although the Frama-C static analyzer performs quite well on s-bAbI, it should be notes that the scores shown are on a 'sound' subset of s-bAbI. On the full dataset even Frama-C's score drops to a 0.85 F1, while GGNN's score remain the same (0.99).

*Summary: Classifying code into vulnerable or not is hard on real-world (noisy) data as opposed to synthetic (clean) data, for both static analyzers and ML/AI approaches alike.*

### B. Encoding: Code-as-graph vs. Others

For the synthetic Juliet dataset (Table 6(b)), GGNN performs better than RNN and CNN (True for both published as well as reproduced results). One reason this might happen, and this was also our hypothesis, is because more meaningful information about code syntax and semantics is preserved in a code-as-graph representation. A GGNN which is able to operate on graph structures is able to then propagate and learn differentiating signals about how the code structure of a vulnerable program differs from that of a healthy one.

|  | F1 | Precision | Recall |
|---|---|---|---|
| Clang_sa | 0.85 | 0.99 | 0.74 |
| Cppcheck | 0.77 | 0.96 | 0.64 |
| Frama-c | 0.98 | 1 | 0.97 |
| Memory Network | 0.91 | 0.95 | 0.86 |
| GGNN | 0.99 | 0.99 | 0.99 |

(a) S-bAbI (sound)

|  | F1 | PR AUC |
|---|---|---|
| Clang_sa | 0.08/0.45 | -- |
| Cppcheck | 0.02/0.05 | -- |
| Frama-c | 0.16 | -- |
| Flawfinder | 0.36 | -- |
| ClassicML (BoW + RF) | 0.78 | 0.89 |
| RNN | 0.8 | 0.9 |
| CNN | 0.84 | 0.94 |
| GGNN | 0.87 | 0.96 |

(b) Juliet

|  | F1 | PR AUC | Draper Subset |
|---|---|---|---|
| ClassicML (BoW + RF) | 0.5 | 0.46 | Debian + Github |
| RNN | 0.53 | 0.46 | Debian + Github |
| CNN | 0.54 | 0.46 | Debian + Github |
| CNN + RF | 0.56 | 0.52 | Debian + Github |
| CNN (reproduced) | 0.53 | 0.44 | Juliet + Debian + Github |
| GGNN | 0.5 | 0.4 | Juliet + Debian + Github |

(c) Draper

Fig. 6. Source-code vulnerability classification performance across 3 datasets: (a) s-bAbI (sound subset), (b) Juliet, and (c) Draper (== Juliet + Debian + Github). Comparison of our GGNN-over-AI4VA approach against static analyzers (Clang_sa, Cppcheck, Frama-C and Flawfinder), classical ML and Deep Learning approaches. Different metrics are shown in different tables based upon the published work from where the competitor scores are borrowed. Legend: PR AUC: Are under the Precision-Recall curve, BoW: Bag-of-Words, RF: Random Forest, Clang_sa: Clang Static Analyzer. Two F1 values are shown for both Cppcheck and Clang_sa static analyzers based upon different values reported by [36] and [31].

Verification of this hypothesis, in terms of interpreting what code structure (nodes and edges in the code graph) influences the classifiers decision, is part of our planned future work.

In case of the Draper dataset, however, we cannot directly compare GGNN scores to those of CNN, since the published results [31] are for two Draper subsets- (i) Juliet-only, and (ii) Github + Debian, instead of all of them combined (i.e. full Draper). When we reproduce[3] the CNN results on the full Draper dataset (last two rows of Table 6(c)), we see slightly lower scores than for its two subsets individually. However, in this case the CNN edges past GGNN (the latter having been modeled only for CWE-119 and CWE-120 vulnerabilities, but better performance not expected with further modeling). There are a few things of note here.

1) The fact that scores for the full Draper dataset (CNN F1=0.53) are slightly lower than those of its two constituent subsets (CNN F1=0.84,0.54) might indicate a problem in combining the differentiating features of synthetic and real-world code.

---
[3]When we reproduced the CNN+RF model on the full Draper dataset, we got much lower scores (F1=0.35) than the ones published for the two Draper subsets (F1=0.82,0.56). We have contacted the authors and are awaiting resolution.

2) The mismatch of a code-as-photo encoding may also be noticed in Table 6(c), where learning with the neural features alone (CNN over pure 'code pixel signals') performed worse than when combined with a RandomForest classifier (CNN + RF) as in [31]. Similarly for RNN + RF, omitted in the table.
3) The scores for the Draper dataset however are so low that a comparison is effectively meaningless. The fact that none of ClassicML, CNN, and GGNN are able to perform much better than a coin-toss may indicate a potential problem with the dataset, in terms of its ability to provide differentiating features (noise vs. signal). The noise can perhaps be decreased by operating on slices [22]. Exploring GGNN effectiveness over CPGs of program slices is another of our future work items.
4) On the other hand, low classification scores across the board may also be symptomatic of the way ground truth labels are collected for Draper, based upon the outputs of static analyzers. We are also looking into alternate sources for generating cleaner data with more trustworthy labels.

*Summary: Treating code as a graph (e.g. AST, CFG, PDG) is beneficial than treating code as a photo or a linear sequence, from an automated learning perspective.*

## C. Classifier's Vulnerability Scope: Generic vs. Specific

sec-pervuln) Since Draper was a particularly challenging dataset, we explored different ways to improve classification performance over it. One of the configurations we tried was to create per-vulnerability models trained only with examples containing the same vulnerability type, instead of a global model trained on all vulnerability types together. We saw promising results in terms of a 52% - 93% increase in accuracy across per-vulnerability models, when using a relatively balanced dataset with a 2:1 non-vulnerable:vulnerable sample ratio. In this setting, instead of classifying a test sample as being vulnerable or not, the methodology would be to run it past individual models and classify it as being susceptible or not to a particular CWE.

*Summary: Learning vulnerability-specific models is easier than a global model across all vulnerabilities.*

## V. RELATED WORK

Traditional approaches to software vulnerability detection can be largely grouped into the categories of static analyses and dynamic analyses [16]. Static analysis tools, such as the Clang static analyzer [2] and Infer [4], typically build a model of program states and reason over all possible behaviors that might happen in the real execution. However, since there are too many possible executions and behaviors, it's usually infeasible to consider all possibilities. Therefore, static analyses in practice usually abstract away some details, lose information and hence produce false positives. For rule-based static analyses, such as linters [1] and taint analysis tools [3], the results depend on the coverage of the defect types and the quality of the rules. It's possible that false negatives can be observed. By contrast, dynamic analyses execute the program and observe the execution behaviors. Testing, symbolic execution, concolic testing and fuzzing [5]–[8], [13], [35] are commonly used for this purpose. While it can concretely expose the defects in the execution, it requires a precise input that can drive the execution to the places of interests. Unfortunately, it's usually challenging to prepare inputs that achieve either good program coverages or satisfying the particular path conditions.

To help alleviate some of these issues with traditional approaches we look towards ML/AI approaches. Since the latter have improved the state of the art in NLP tasks and since code also has a language-like 'naturalness' to it [10], perhaps even more structured than natural language, we hypothesized it would be worth exploring learning-based approaches to source code vulnerability analysis which respect its natural structure.

We base our work primarily on Code Property Graphs (CPG) introduced by Yamaguchi et al. [39], and Gated Graph Neural Networks proposed by Li et al. [21]. The former proposed the idea of using manually created templates for software vulnerabilities for graph traversals on a union (==CPG) of the code's AST, CFG and PDG graphs. And the latter optimizes Graph Neural Networks [34] to extract features from graph-structured inputs, and use it to solve tasks such as logic reasoning, path finding, and program invariant inferencing. We combine both of these approaches to automatically learn signatures of vulnerabilities in source code.

From a security perspective, several machine learning approaches have been developed to learn differentiating signals from explicit code features for software vulnerability detection or code quality evaluation. These tend to learn from source code features such as number of lines or conditional statements, use of sensitive library functions or system calls, call-stack depth, complexity measures like cyclomatic or Halstead complexity, etc. [9], [18], [19], [26], [33], [37]. This can be augmented with meta features such as commit messages and bug reports [32], [40]. Alternatively, deep learning approaches automatically extract signals from code by treating it as a photo or a linear sequence [14], [20], [22], [23], [30], [31]. We believe more meaningful signals are preserved in a graph-level encoding of source code and thus experiment with a code-as-a-graph representation.

From an encoding perspective, recent work has also looked into deep learning over code-as-graph as in [11], [12], [38]. [38] uses Graph Neural Network based auto-encoders for unsupervised learning over source code ASTs to automatically cluster Java classes into different categories- business logic, interface, and utility classes. [11] also uses Gated Graph Neural Networks, and targets learning lower / fundamental levels of program structures such as predicting variable names given its usage, or correct variable selection given a program location. Similarly, code2vec [12] uses a collection of path-contexts (AST paths + leaf-node values) to encode source code, and use an attention-based neural network to learn appropriate names for a function from its code. On the other hand, we operate at a macroscopic level of automatically learning vulnerability templates from source code graphs.

For an exhaustive survey of the variants and use-cases of machine learning over source code, we refer the reader to a 2018 paper by Allamanis et al. [10].

## VI. CONCLUSION AND FUTURE WORK

In this work we explored the possibility of automatically extracting signatures of software vulnerabilities, in terms of node connectivity templates over a graph-level representation of source code. We developed an AI4VA pipeline to classify C functions as being vulnerable or not, by converting them into vectorized graph representations while preserving their structural and semantic information. We trained a state-of-the-art Gated Graph Neural Network to learn to differentiate the properties of code graphs of vulnerable functions from their healthy counterparts. Our model outperformed static analyzers, classical machine learning, and deep learning approaches in two of the three datasets we experimented with. On the third dataset, the performance across the board was not much better than a coin-toss, prompting us to look into cleaner real-world datasets with more trustworthy ground truth labels as part of ongoing work. We also uncovered issues with combining the differentiating features of synthetic and real-world code, leading us to suspect if transfer learning from a high quality synthetic dataset to real world code would be feasible. Our results indicate a superiority of a code-as-graph approach to vulnerability detection as compared to treating code as a photo or a linear sequence as in previous approaches. We also observed, probably as expected, that vulnerability-specific models performed better than a global model across all vulnerabilities.

In addition to looking into alternate means of collecting a cleaner real-world dataset, other threads of our ongoing work include:

1) Pinpointing the vulnerable locations in the code (nodes in the CPG), as well as interpreting the classifier's decision process by uncovering the vulnerability templates (node connectivity in the CPGs) being learnt by it during training. We have started looking into tracking back from 'influencer' nodes, in terms of which other nodes, and following which particular edges, did these nodes accumulate their discerning signals.
2) Manually inspecting the incorrectly classified functions to understand reasons for mistakes (False Positives / Negatives). When we manually inspected the functions which were incorrectly classified in the Draper dataset, we found it challenging to even manually tag them as being vulnerable or not. We have thus redirected our focus to inspecting functions on simpler datasets–Juliet, s-bAbI, to gain any insights on how the model could be further improved.
3) Training on graphs of program slices instead of full source code to reduce noise in data [22].